\documentclass{article}
\usepackage{spconf,amsmath,epsfig}

\usepackage[]{graphicx}
\graphicspath{{Figures/}}
\usepackage{caption}
\usepackage{subfig}
\usepackage{amsmath}
\usepackage{amssymb}
\usepackage{epsfig}
\usepackage{cite}
\usepackage{color}
\usepackage{balance}
\usepackage{amsmath}
\usepackage{amsfonts} 
\usepackage{graphicx}
\usepackage{pdfpages}
\usepackage[T1]{fontenc} 
\usepackage{array}
\usepackage{url}

\usepackage{color}
\usepackage{wrapfig}
\usepackage{lipsum}
\usepackage[hidelinks]{hyperref}
\usepackage{multirow}
\usepackage{tabularx}
\usepackage{gensymb} 
\usepackage{breqn}
\usepackage{array}
\usepackage{booktabs}

\title{A Data-driven Approach to Detecting Precipitation \\from Meteorological Sensor Data}

\name{Shilpa Manandhar$^{1}$, Soumyabrata Dev$^{2}$, Yee Hui Lee$^{1}$, Yu Song Meng$^{3}$ and Stefan Winkler$^{4}$
\thanks{This research is funded by the Defence Science and Technology Agency (DSTA), Singapore.}
\thanks{Send correspondence to S.\ Manandhar: \url{shilpa005@e.ntu.edu.sg}}
}
\address{
	$^{1}$~School of Electrical and Electronic Engineering, Nanyang Technological University, Singapore \\
	$^{2}$~ADAPT Centre, School of Computer Science and Statistics, Trinity College Dublin, Ireland \\     
    $^{3}$~National Metrology Centre, Agency for Science, Technology and Research (A$^{*}$STAR), Singapore \\
    $^{4}$~Advanced Digital Sciences Center (ADSC), University of Illinois at Urbana-Champaign, Singapore
}

\begin{document}
\maketitle
\begin{abstract}
Precipitation is dependent on a myriad of atmospheric conditions. In this paper, we study how certain atmospheric parameters impact the occurrence of rainfall. We propose a data-driven, machine-learning based methodology to detect precipitation using various meteorological sensor data. Our approach achieves a true detection rate of 87.4\% and a moderately low false alarm rate of 32.2\%. 
\end{abstract}


\section{Introduction}
\label{sec:intro}
Precipitation initiation is a dynamic process that is influenced by many weather variables, location, and seasons. In general terms, as a parcel of air containing water vapor rises in the atmosphere, it will reach a certain height at which the temperature drops below the dew point and the air becomes saturated. Any small excess of water vapor beyond this saturation point will cause the excess amount of vapor to condense into liquid water or ice, forming clouds~\cite{book}. Further ascent of water vapor can lead to the growth of clouds, which may finally precipitate. However, predicting rainfall from the behavior of different weather parameters is challenging. 

The research community has shown growing interest in rainfall prediction over the past few years. Recent publications \cite{Yibin(a),Benevides} have discussed using precipitable water vapor (PWV) content derived from Global Positioning System (GPS) signal delay to predict the rainfall. We have also used GPS-derived PWV values for rain prediction \cite{IGARSS_2016} and have employed sky cameras \cite{Dev2016GRSM} to detect the onset of precipitation~\cite{rainonset}. However, the water vapor content of the atmosphere -- albeit a good indicator of rain -- is not sufficient to predict rain with high accuracy. Other researchers \cite{Junbo,sharifi} have suggested using other meteorological parameters. In ~\cite{manandhar2018systematic}, we have provided a systematic analysis of the various weather parameters for rainfall detection.

In this paper, we use various surface weather parameters along with the water vapor content derived from GPS, and implement a machine learning based technique to classify \emph{rain} and \emph{no rain} observations. In the following sections, we describe the dataset used in this paper and present the proposed algorithmic approach. Then we discuss the experiments and test results. Finally, we conclude the paper with future directions. The source code of all simulations in this paper is available online.\footnote{~\url{https://github.com/shilpa-manandhar/precipitation-detection}}

\section{Features for Rainfall Classification}
In this section, we describe the different variables, including surface weather parameters, total column water vapor content, and seasonal/diurnal characteristics, which are later used for rainfall classification.  

\subsection{Surface Weather Parameters}
Surface weather parameters recorded by a weather station (Davis Instruments 7440 Weather Vantage Pro II) with tipping bucket rain gauge are used in this study. The weather station is located at Nanyang Technological University (NTU), (1.3$^{\circ}$N, 103.68$^{\circ}$E). The following weather station measurements are used for this paper:
\begin{itemize}
\setlength\itemsep{1pt}
\item Surface temperature ($^{\circ}$C), 
\item Relative humidity (RH) (\%), 
\item Dew point ($^{\circ}$C),
\item Solar irradiance (W/m$^{2}$),
\item Rainfall rate (mm/hr). 
\end{itemize}
All the recorded weather parameters have a temporal resolution of 1 minute. 

\begin{figure*}[htb]
\centering
\includegraphics[width=0.65\textwidth]{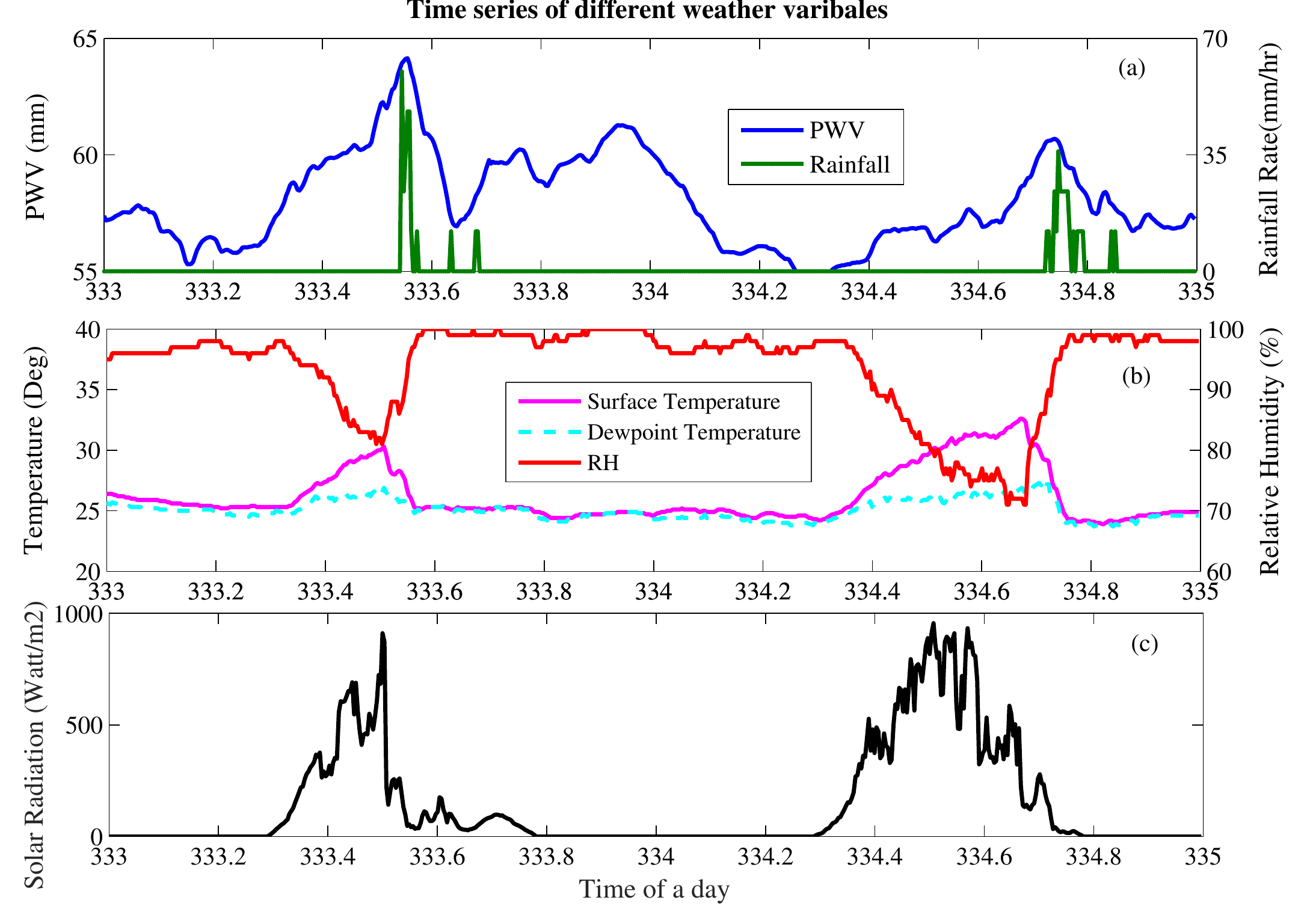}
\caption{Time series of different weather parameters. (a) PWV and Rainfall rate; (b) Surface Temperature, Dew point, and Relative humidity; (c) Solar irradiance. The horizontal axis for all subplots represents the time in day of the year (\textit{DoY}). For example, 334.8 indicates November 30 at 19:20.
\label{TimeSeries}}
\end{figure*} 

\subsection{GPS Derived Water Vapor Content}
In addition to the various surface weather parameters, precipitable water vapor (PWV) values derived from GPS signal delays are used as an additional important feature for the classification. This section provides a brief overview of the derivation of PWV values from GPS signal delays. 

PWV values (in mm) are calculated using the zenith wet delay (ZWD), \textit{$\delta$L$_w^{o}$}, incurred by the GPS signals as follows: 
\begin{equation}
	\mbox{PWV}=\frac{PI \cdot \delta L_w^{o}}{\rho_l},
    \label{eq1}
\end{equation}     
where $\rho_{l}$ is the density of liquid water (1000 kg$/m^{3}$), and \textit{PI} is the dimensionless factor determined by \cite{shilpaPI}:
\begin{dmath}
	PI=[-\text{sgn}(L_{a})\cdot 1.7\cdot 10^{-5} |L_{a}|^{h_{fac}}-0.0001]\cdot \\ 
    \cos \frac{2\pi(DoY-28)}{365.25}+
    0.165-1.7\cdot 10^{-5}|L_{a}|^{1.65}+f,
    \label{eq2}
\end{dmath}
where $L_a$ is the latitude, \textit{DoY} is day-of-year, $h_{fac}$ is either 1.48 or 1.25 for stations in the Northern or Southern hemisphere, respectively.  $f=-2.38\cdot 10^{-6}H$, where $H$ is the station altitude above sea level, can be ignored for $H<1000m$. 

For this paper, the ZWD values for an IGS GPS station located at NTU (station ID: NTUS) are processed using GIPSY OASIS software and recommended scripts \cite{GIPSY}. PWV values are then calculated for NTUS using Eqs.~\ref{eq1}-\ref{eq2}, with $L_a= 1.34$, $h_{fac}= 1.48$, $H=78m$. The calculated PWV values have a temporal resolution of 5 minutes. 

\subsection{Seasonal and Diurnal Features}
Singapore experiences four different seasons:
\begin{itemize}
\setlength\itemsep{1pt}
\item North-East Monsoon (NE) during November-March,
\item First-Inter Monsoon (FI) during April-May,
\item South-West Monsoon (SW) during June-October,
\item Second-Inter Monsoon (SI) during October-November.
\end{itemize}
The period of these seasons vary a little from year to year, which is updated in a yearly weather report provided by Singapore's National Environment Agency (NEA) \cite{NEA}. 

The rainfall pattern in Singapore is heavily influenced by different seasons. Most of the rain is experienced in NE and SW monsoon seasons. Since seasons play an important role in rainfall, we include day-of-year (\textit{DoY}) as a feature for rainfall classification. Furthermore, rainfall occurrences in tropical regions like Singapore also show clear diurnal characteristics. Heavy convective rainfalls are generally experienced during the late afternoon in the NE monsoon in Singapore. Thus time-of-day (\textit{ToD}) is also included as a feature.  

\subsection{Time Series Example}
In this section, we show time series data to illustrate the importance of all the features for rain classification. Fig.~\ref{TimeSeries} shows the time series of weather parameters over two consecutive days in 2010. Weather parameters are sampled at 5-minute intervals to match the GPS PWV timings.

The different features show interesting changes with rain. PWV values tend to increase before the rain. Surface temperature decreases and matches the dew point temperature during the rain. Relative humidity increases and reaches almost 100 \% when it rains and RH is also generally higher in the night time. Similar fluctuations can be observed in the solar irradiance values. For Singapore, clear sky solar radiation is around 1000 W/m$^{2}$ in the daytime~\cite{ClearSkyPIERS}. In Fig.~\ref{TimeSeries}(c) we can see the drop in the solar radiation  before  rain, likely due to the buildup of clouds. The example also highlights the diurnal variations of rain and these weather parameters, as discussed in the previous section.

\section{Experiments}
\subsection{Database}
In this paper, three years (2010-2012) of data are used for the experiments. The data from 2010-2011 are used for training and testing the algorithm. For further assessment of the performance, the trained algorithm is also tested on data from a separate year (2012). 

\subsection{Dataset Imbalance}
For any classification algorithm it is very important to train a model properly, and thus the training data should be chosen wisely. In this paper, we consider $7$ features -- \textit{temperature}, \textit{dew point}, \textit{relative humidity}, \textit{solar irradiance}, \textit{PWV}, \textit{ToD}, and \textit{DoY}, which are used for binary classification of rain. Each of these features are used with a temporal resolution of 5 min. If a dataset of 1 year (365 days) is taken as the training database, it includes a total of 365*288 data points. Out of these 105,120 data points, there are far fewer data points with rain than without, because rain is relatively rare event.

For a year's (2010) data, the ratio of data points with rain (\textit{minority cases}) to the data points without rain (\textit{majority cases}) is 1:70, which indicates that the database  is highly imbalanced with respect to rain events. Consequently, the traditional way of separating a database with say 70\% of training size and 30\% of test size might result in a biased model, which is dominated by the characteristics of the majority database. Instead, we employ random downsampling techniques to make the training dataset balanced \cite{RandomDownSampling}.

Random downsampling is one of the techniques used to overcome the problem of imbalanced databases. In this method, while forming the training data set, all the cases from the minority scenario are taken into consideration. Then the cases from majority scenario are randomly chosen such that the minority to majority ratio is balanced. There is a general practice to make the ratio 1:1, but other ratios can also be considered \cite{RandomDownSampling}.  

\subsection{Training and Testing}
A certain percentage of data points from two years (2010-2011) of the database is taken randomly as a training set and the remaining data as the test set. The training dataset is then balanced by performing random downsampling to obtain a minority to majority ratio of 1:1. The balanced training data is then used to train the model using a Support Vector Machine (SVM). The model is trained for different training data sizes. Since the training data is selected randomly, for each training data size, the experiment is performed 100 times and the average values of the evaluation metrics are calculated.

\subsection{Evaluation Metrics}
There are different evaluation metrics that can be used to analyze the results. One should choose a suitable evaluation metric that best fits the scenario. For the study of rain, it is important to see how well the rainfall is predicted and how often the methodology makes false predictions. Therefore, the results are generally expressed in terms of true detection and false alarm rates \cite{Yibin(a), Benevides}.
From the confusion matrix, the true positive, true negative, false positive and false negative samples are represented by $TP$, $TN$, $FP$ and $FN$ respectively. We report the True Detection (TD) and False Alarm (FA) rates, which are defined as follows:
\begin{equation*}
TD = TP/(TP+FN),
\end{equation*}
\begin{equation*}
FA = FP/(TN+FP).
\end{equation*}

\section{Results}

\begin{figure}[ht]
\begin{center}
\includegraphics[width=0.45\textwidth]{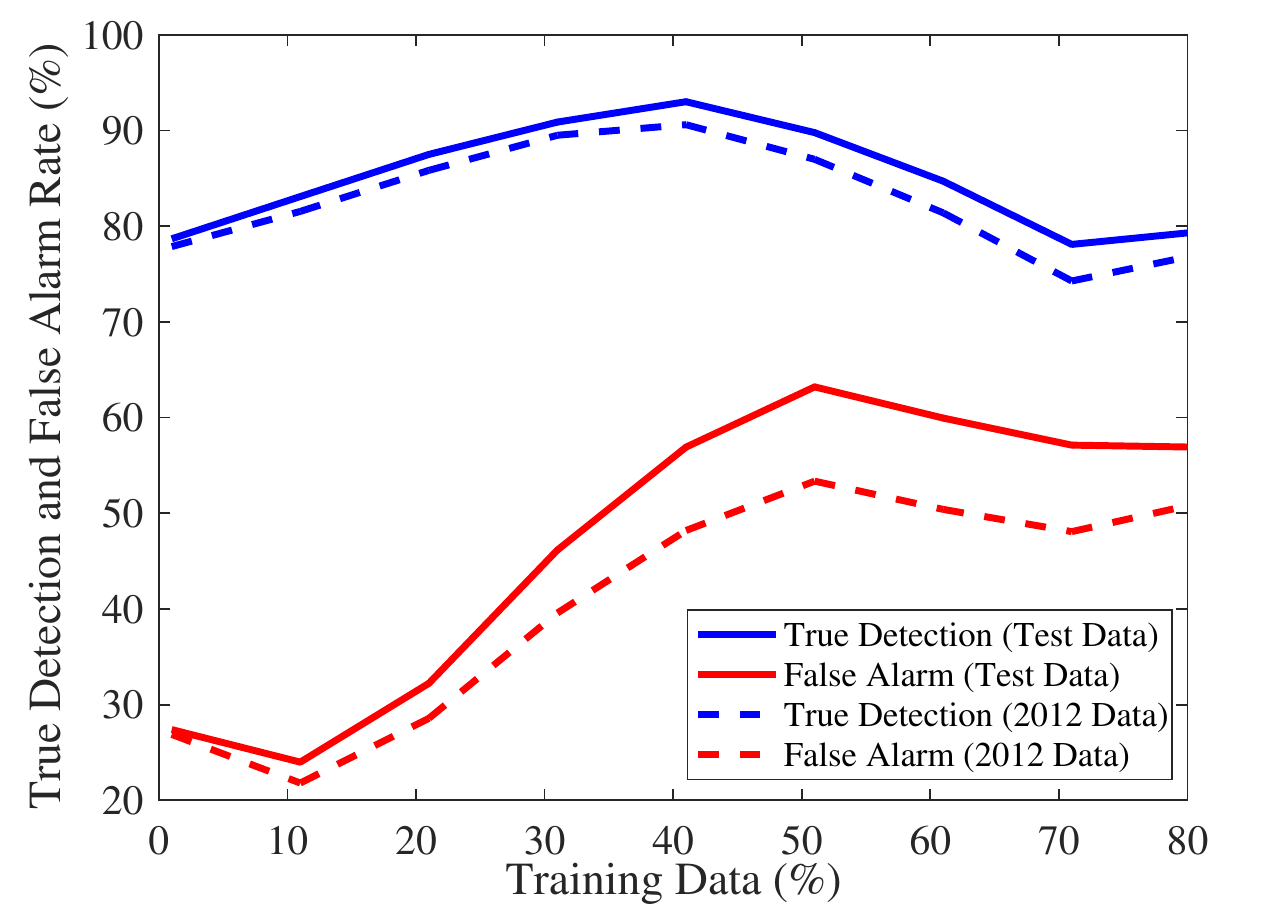}
\caption{True Detection and False Alarm rates for test data and data from 2012.
\label{fig:meas-rec}}
\end{center}
\end{figure} 

Fig.~\ref{fig:meas-rec} shows the average TD and FA values at varying \% of training data size. The model reaches the highest TD at around 40\% of training data size, but the FA is also quite high at this point. Around a training data size of 20\% or lower the TD is good and the FA values are also lower. Similar results were obtained when the trained model was tested on data from a separate year (2012).

Table~\ref{tab:result} reports the TD and FA rates at 20\% of training data size and remaining observations as testing data set with and without downsampling. The results with downsampling are better as compared to those without. We achieve a high true detection rate of 87.4\% and 85.8\% on the test data and data from 2012 respectively. Similarly, the false alarm rate is 32.2\% and 28.5\% for test data and data from 2012 respectively. In the literature~\cite{Benevides,Yibin(a)}, a true detection rate of 80\% and a false alarm rate of 60\% has been reported for rainfall prediction using PWV data. The results presented in this paper show a significant improvement in the false alarm rate reported. Therefore, our approach is able to achieve a better performance for rain detection.

\begin{table}[ht]
\centering
\caption{True Detection (TD) and False Alarm (FA) rates in \% at 20\% training data size.}
\begin{tabular}{*5c}
\toprule
Dataset &  \multicolumn{2}{c}{\shortstack{Without \\ downsampling}} & \multicolumn{2}{c}{\shortstack{With \\ downsampling}} \\
\midrule
{}                & TD & FA  & TD & FA \\
Test Data         &  82.5  & 44.8    & 87.4   & 32.2\\
2012 Data   &  82.9  &  42.4   & 85.8   & 28.5\\
\bottomrule
\end{tabular}
\label{tab:result}
\end{table}

\section{Conclusions \& Future Work}
\label{sec:conc}
In this paper, a machine-learning based framework to detect precipitation from regular surface meteorological parameters and GPS derived PWV has been presented. Our proposed method has a high true detection rate and moderately low false alarm rate, as demonstrated using weather data from the tropics. In the future, we plan to use set-theory based techniques~\cite{dev2017rough} to analyze the impact of these various features on precipitation. We also plan to study methodologies to further reduce false alarms. We will also explore other techniques to counter the effects of an unbalanced dataset with rare events of interest.

\balance

\end{document}